\documentclass[12pt]{iopart}
\usepackage{graphicx}
\usepackage{hyperref}
\usepackage{amssymb}
\usepackage[numbers,sort&compress]{natbib}
\usepackage{amsmath}
\usepackage{appendix}
\usepackage{color}

\hypersetup{hypertex=true,
colorlinks=true,
linkcolor=blue,
urlcolor = blue,
citecolor=blue}
\begin{document}

\title{Dynamics of non-Hermitian Floquet Wannier-Stark system}
\author{H. P. Zhang{$^{1}$}, K. L. Zhang{$^{2}$} and Z. Song{$^{1,\dagger}$}}

\begin{abstract}
We study the dynamics of the non-Hermitian Floquet Wannier-Stark system in
the framework of the tight-binding approximation, where the hopping strength
is a periodic function of time with Floquet frequency $\mathcal{\omega }$.
It is shown that the energy level of the instantaneous Hamiltonian is still
equally spaced and independent of time $t $ and the Hermiticity of the
hopping term. In the case of off resonance, the dynamics are still periodic,
while the occupied energy levels spread out at the resonance, exhibiting $%
t^{z}$ behavior. Analytical analysis and numerical simulation show that the
level-spreading dynamics for real and complex hopping strengths exhibit
distinct behaviors and are well described by the dynamical exponents $z=1$
and $z=1/2$, respectively.
\end{abstract}

\address{{$^{1}$} School of Physics, Nankai University, Tianjin 300071,
China} 
\address{{$^{2}$}School of Physics and Optoelectronic Engineering,
Foshan University, Foshan,\newline \phantom{\,} 528000, China} \ead{{$^{%
\dagger}$}songtc@nankai.edu.cn}

\vspace{10pt} 

\maketitle

\section{Introduction}

\label{Introduction}The Hermitian and static system is the main study object
of traditional quantum mechanics, preserving the reality and conservation of
energy. When these two constraints are violated, the properties of the
material could be drastically modified, leading to new phases and phenomena 
\cite{Moiseyev2011,Tannor2007} without analogs in their static or Hermitian
counterparts. {Floquet engineering has facilitated the exploration of a
diverse range of dynamic, topological, and transport phenomena across
various physical contexts, including platforms involving ultracold atoms 
\cite{Schweizer2019,Wintersperger2020,Wintersperger2020a,Liu2022}. More
recently, the Floquet approach has been employed to engineer topological
phases in non-Hermitian systems. Importantly, it was discovered that the
interplay between time-periodic driving fields and the presence of gain,
loss, or nonreciprocal effects can lead to the emergence of topological
phases exclusive to non-Hermitian Floquet systems \cite%
{PhysRevB.98.205417,PhysRevLett.102.065703,LeonMontiel2018,Li2019,PhysRevA.102.062201,Blose2020,Zhou2020,Zhang2020,Xiao2020,Ageev2021}
. }

Motivated by the above investigations, the aim of the present paper is to
examine how non-Hermiticity and periodic driving impact a Stark ladder
system. Specifically, we study the dynamics of the non-Hermitian Floquet
Wannier-Stark system in the framework of the tight-binding approximation. A
conventional Wannier-Stark ladder consists of quantized equidistant energy
levels, which separated by the Bloch frequency and emerge in a periodic
Hermitian system under a linear potential \cite%
{Glueck2002,Glueck1999,Kolovsky2002}. Based on this framework, the Bloch
oscillations in cold atoms within the optical lattice have also been
theoretically investigated \cite{Kolovsky2003,Kolovsky2009,Kolovsky2010}.

{{Recent works \cite{Longhi2015,Graefe2016,Longhi2017,Qin2020} show that
non-Hermitian hopping terms, including asymmtric and complex strengths,
remain the reality of the Wannier-Stark ladder. On the other hand, it has
proposed that a time-dependent\ hopping strength for a Bloch system can be
realized by\ dynamically modulated waveguide arrays \cite{Zhan2022}.}}

In this work, we introduce non-Hermiticity and periodic driving by focusing
on cosinusoidal time-dependent\ hopping strength with frequency$\mathcal{\
\omega }$ and complex amplitude. The main reason for choosing such systems
is that the energy level of the instantaneous Hamiltonian is still equally
spaced with $\mathcal{\omega }_{0}$\ and\ independent of time $t$ and the
Hermiticity of the hopping strength. This allows us to investigate the
dynamics based on analytical solutions. We find that, in the case of off
resonance, the dynamics are still periodic, while the occupied energy levels
spread out\ at the resonance, exhibiting $t^{z}$ behavior. Analytical
analysis and numerical simulation show that the level-spreading dynamics for
real and complex hopping strengths exhibit distinct behaviors and are well
described by the dynamical exponents $z=1$ and $z=1/2$, corresponding to
superdiffusion and diffusion, respectively \cite%
{Zhao2006,metzler1999anomalous,Bouchaud1990}. These findings may help us
deepen our understanding of the interplay between non-Hermiticity and
periodic driving impacts on the dynamics of the system. In addition, we
propose a scheme to demonstrate the results by wavepacket dynamics in a $2$D
square lattice.

This paper is organized as follows. In Sec. \ref{Model and generalized
Wannier-Stark ladder}, we introduce a non-Hermitian time-dependent
tight-binding model supporting a Wannier-Stark ladder and present the
analytical eigenstates of the instantaneous Hamiltonian. In Sec. \ref{Bloch
oscillation in non-Hermitian system}, we focus on the dynamics of the static
system, while the Floquet system at resonance in Sec. \ref{Floquet-resonance
dynamics}. Sec. \ref{Dynamical exponents} describes the application of the
solutions to a specific initial state. In Sec. \ref{2D simulator via
wavepacket dynamics}, a scheme for the simulator of our results is proposed
in the framework of wavepacket dynamics in a $2$D square lattice.\ Finally,
we summarize our results in Sec. \ref{Conclusion}.

\begin{figure}[t]
\centering
\includegraphics[width=0.85\textwidth]{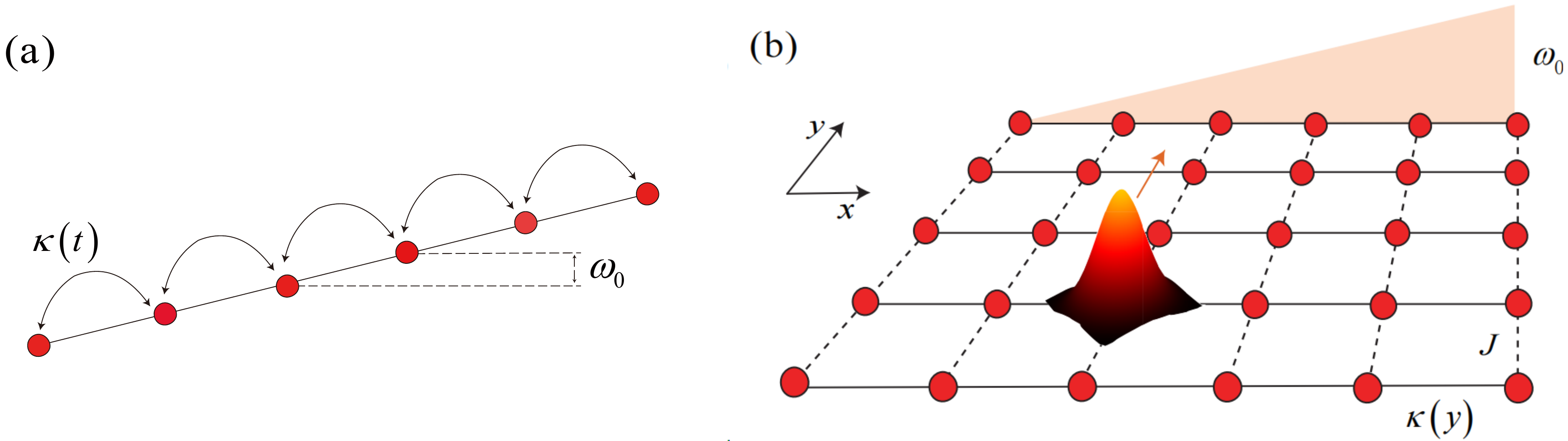}
\caption{(a) Schematic illustrations of the Hamiltonian in Eq. (\protect\ref%
{e1}), which represents a tight-binding chain with a time dependence on the
hopping strength between two adjacent sites. The on-site potential is tilted
with slope\ $\protect\omega _{0}$. It is shown that such a system supports a
Stark-ladder with time-independent energy-level spacing. (b) Schematic
illustrations of the Hamiltonian in Eq. (\protect\ref{H2D}), which is a
tight-binding model of a square lattice with time-independent system
parameters. The contributions of hopping strength and on-site potential can
be used to simulate the dynamics of system (a).}
\label{fig1}
\end{figure}

\section{Model and generalized Wannier-Stark ladder}

\label{Model and generalized Wannier-Stark ladder}

We start with the tight-binding model with the Hamiltonian 
\begin{equation}
H=\kappa (t)\sum\limits_{n=-\infty }^{+\infty }(|n\rangle \langle n+1|+\text{
\textrm{H.c}})+\mathcal{\omega }_{0}\sum\limits_{n=-\infty }^{+\infty
}n|n\rangle \langle n|,  \label{e1}
\end{equation}%
where $|n\rangle $ denotes a site state describing the Wannier state
localized on the $n$th period of the potential. Here, $\kappa (t)$ is the
time-dependent hopping strength, while $\mathcal{\omega }_{0}$\ is the slope
of the linear potential from a static force. Fig. \ref{fig1}(a) shows a
schematic of the system. In our study, $\kappa (t)$\ can be real and
complex. { In order to solve the time-dependent Hamiltonian, we
start with the solution for the case where $\kappa $ is an arbitrary complex
constant. We then discuss the time-dependent hopping strength $\kappa (t)$
at the end of the Section \ref{Bloch oscillation in non-Hermitian system}. }%
The non-Hermitian Hamiltonians with complex hoppings have been realized in
the coupled waveguides \cite{Zhang2019}, optical fibres \cite{Bergman2021}
and electrical circuit resonators \cite{Choi2018}.

{ It is well known that when the hopping constant $\kappa $ is
real, the Hamiltonian ${H}$\ in momentum space, based on the Bloch basis
representation, can be exactly diagonalized. Consequently, the energy levels
of ${H}$ are equidistant, which supports the periodic dynamics \cite%
{Hartmann2004,bloch1928elektron,1934}.} This phenomenon has been observed in
systems of semiconductor superlattices and ultracold atoms \cite%
{Waschke1993,BenDahan1996,Wilkinson1996,Anderson1998}. {Now, we
extend the conclusion to the case with arbitrary $\kappa $ through an
alternative approach.} The eigenstate of ${\left\vert \psi _{m}\right\rangle 
}$\ can always be written in the form 
\begin{equation}
{\left\vert \psi _{m}\right\rangle =\sum_{n}c_{nm}|n\rangle ,}
\end{equation}%
satisfying the Schr\"{o}dinger equation 
\begin{equation}
{H\left\vert \psi _{m}\right\rangle =E_{m}\left\vert \psi _{m}\right\rangle .%
}
\end{equation}%
{ The eigenenergy $E_{m}$ can be obtained by the equation 
\begin{equation}
E_{m}={\kappa }\frac{c_{n+1,m}+c_{n-1,m}}{c_{nm}}+n\omega _{0},
\end{equation}%
which accords with the recurrence relation of the Bessel function {$J_{n}$}
of argument $-2\kappa /\omega _{0}$, i.e., 
\begin{equation}
m\omega _{0}={\kappa }\frac{J_{n-m+1}(-\frac{2\kappa }{\omega _{0}}%
)+J_{n-m-1}(-\frac{2\kappa }{\omega _{0}})}{J_{n-m}(-\frac{2\kappa }{\omega
_{0}})}+n\omega _{0}.
\end{equation}%
Then, we have 
\begin{equation}
E_{m}=m\omega _{0},\ {c_{nm}=J_{n-m}(-\frac{2\kappa }{\omega _{0}}).}
\end{equation}%
} We can see that the eigenenergy ${m\mathcal{\omega }_{0}}$ is independent
of $\kappa $, regardless of whether it is real or complex. This is crucial
for the main conclusion of this work. Although the profile of the eigenstate 
\begin{equation}
\psi _{m}\left( n\right) =\langle n|\psi _{m}\rangle =J_{n-m}(-\frac{2\kappa 
}{\omega _{0}}),
\end{equation}%
in real space is $\kappa $ dependent, and it is always localized. The
inverse participation ratio (IPR) is a simple way to quantify the
localization of a given state. For spatially extended states, the value of
the IPR approaches zero when the system is sufficiently large, whereas it is
finite for localized states regardless of the system size. The IPR of the
eigenstate $\left\vert \psi _{m}\right\rangle $\ is 
\begin{equation}
{\ \text{IPR}_{m}=\frac{\sum\limits_{n}\left\vert \langle n|\psi _{m}\rangle
\right\vert ^{4}}{(\sum\limits_{n}\left\vert \langle n|\psi _{m}\rangle
\right\vert ^{2})^{2}}=\frac{\sum_{n}|J_{n-m}(-\frac{2\kappa }{\omega _{0}}%
)|^{4}}{(\sum_{n}|J_{n-m}(-\frac{2\kappa }{\omega _{0}})|^{2})^{2}}.}
\end{equation}%
Numerical simulations show that IPR$_{m}$\ is independent of $m$\ and $N$\
for a sufficiently large $N$. We have IPR$_{m}=0.418$\ and ${0.534}$ for $%
-2\kappa /\omega _{0}=1.0$\ and $1.0i$, respectively, which indicates that $%
\left\vert \psi _{m}\right\rangle $\ is in a localized state. {
Although it is practically impossible to have a system of infinite size,
when considering a system of finite size where the system size exceeds the
width of {$\left\vert \psi _{m}\right\rangle $}, the energy levels, with the
exception of a few edge levels influenced by finite size effects, remain
equidistant.{\ To demonstrate this point, numerical results for the energy
levels of finite-sized non-Hermitian chains are plotted in Fig. \ref%
{refereefig1}. The figure shows that: (i) the complex energy levels do not
form conjugate pairs; (ii) as the chain size increases, the size of the
Wannier-Stark ladder within the central region also increases. These results
suggest that the existence of a real Wannier-Stark ladder is attributed to
the infinite length of the chain, rather than the pseudo-Hermiticity of the
system.}}

\begin{figure}[tbh]
\centering
\includegraphics[width=0.85\textwidth]{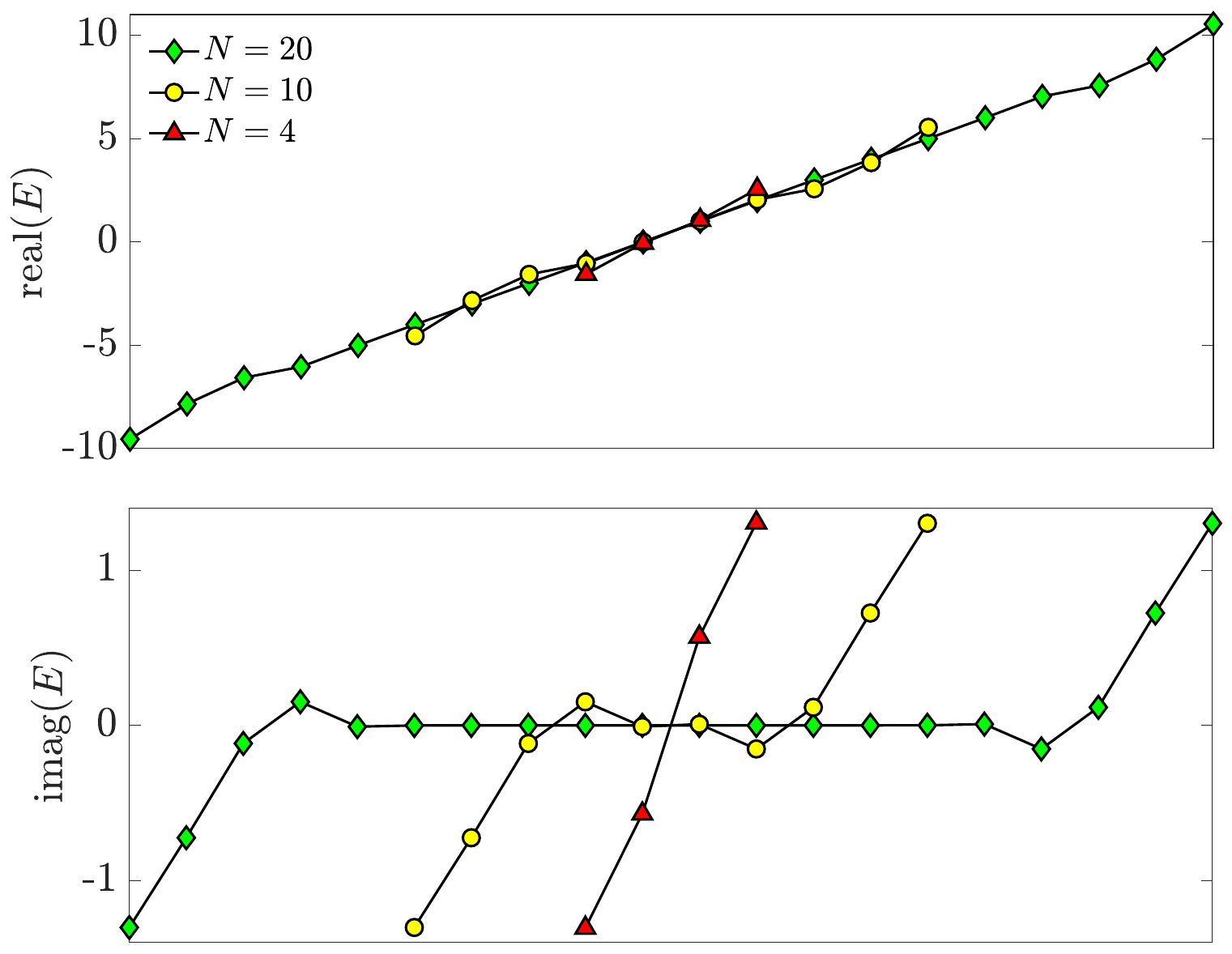}
\caption{Plots of the real and imaginary parts of energy levels
for the system at different sizes. The parameters of the system are $\protect%
\kappa =1+1i$ and $\protect\omega _{0}=1$. One can observe that the real,
equally spaced energy levels emerge from the center of the spectrum as the
system size increases.}
\label{refereefig1}
\end{figure}

In parallel, without loss of generality, we have 
\begin{equation}
\left\vert \varphi _{m}\right\rangle =\sum_{n}J_{n-m}(-\frac{2\kappa ^{\ast
} }{\omega _{0}})|n\rangle ,
\end{equation}
for the equation 
\begin{equation}
H^{\dag }\left\vert \varphi _{m}\right\rangle =m\omega _{0}\left\vert
\varphi _{m}\right\rangle ,
\end{equation}
which establishes the biorthonormal set $\left\{ |\varphi _{m}\rangle ,|\psi
_{n}\rangle \right\} $, satisfying 
\begin{equation}
\langle \varphi _{m}\left\vert \psi _{n}\right\rangle =\sum_{n^{\prime
}}J_{n^{\prime }-m}^{\ast }(-\frac{2\kappa ^{\ast }}{\omega _{0}}
)J_{n^{\prime }-n}(-\frac{2\kappa }{\omega _{0}})=\delta _{mn},
\end{equation}
since the eigenlevels are independent of $\kappa $, without coalescence of
eigenvectors. This is an important basis for the following investigations.

The obtained solution to the non-Hermitian Hamiltonian obeys the
well-established theory of non-Hermitian quantum mechanics \cite%
{Bender1998,Mostafazadeh2002,Mostafazadeh2002a,Mostafazadeh2002b,Bender2007}%
. In the context of non-Hermitian systems with $H\neq H^{\dag }$,
biorthonormal complete eigenstates refer to a set of eigenstates that are
biorthonormal and form a complete basis for the Hilbert space of the system.
The eigenstates of a non-Hermitian Hamiltonian can be complex and may not be
orthogonal to each other. However, if a non-Hermitian Hamiltonian has a
complete set of biorthonormal eigenstates, it means that: There exists a
left eigenstate $\langle \varphi _{n}|$ and a right eigenstate $|\psi
_{n}\rangle $, satisfying 
\begin{equation}
H|\psi _{n}\rangle =E_{n}|\psi _{n}\rangle ,H^{\dag }|\varphi _{n}\rangle
=E_{n}^{\ast }|\varphi _{n}\rangle .
\end{equation}
for each eigenvalue $E_{n}$ such that 
\begin{equation}
\langle \varphi _{n}|\psi _{m}\rangle =\delta _{nm},\sum_{n}|\psi
_{n}\rangle \langle \varphi _{n}|=1.
\end{equation}
The set of right eigenstates $\left\{ |\psi _{n}\rangle \right\} $ and the
set of left eigenstates $\left\{ |\varphi _{n}\rangle \right\} $ together
form a basis for the Hilbert space, allowing any state in the space to be
expanded in terms of these eigenstates.

The concept of biorthonormal complete eigenstates is particularly important
in non-Hermitian quantum mechanics because it plays an analogous role, both
conceptually and computationally, to those in Hermitian quantum systems. For
example, when we calculate the time evolution of a state, we first express
the initial state $|\psi (0)\rangle $\ as a linear combination of the right
eigenstates 
\begin{equation}
|\psi (0)\rangle =\sum_{n}|\psi _{n}\rangle \langle \varphi _{n}|\psi
(0)\rangle
\end{equation}%
by projecting the initial state onto the left eigenstates. The state of the
system at any later time $t$ is given by 
\begin{equation}
|\psi (t)\rangle =\sum_{n}|\psi _{n}\rangle e^{-iE_{n}t}\langle \varphi
_{n}|\psi (0)\rangle =U(t)|\psi (0)\rangle
\end{equation}%
where $U(t)$ is time evolution operator and will be obtained in the
following for {the present Hamiltonian in Eq. (\ref{e1}).}

\begin{figure}[tbh]
\centering
\includegraphics[width=0.9\textwidth]{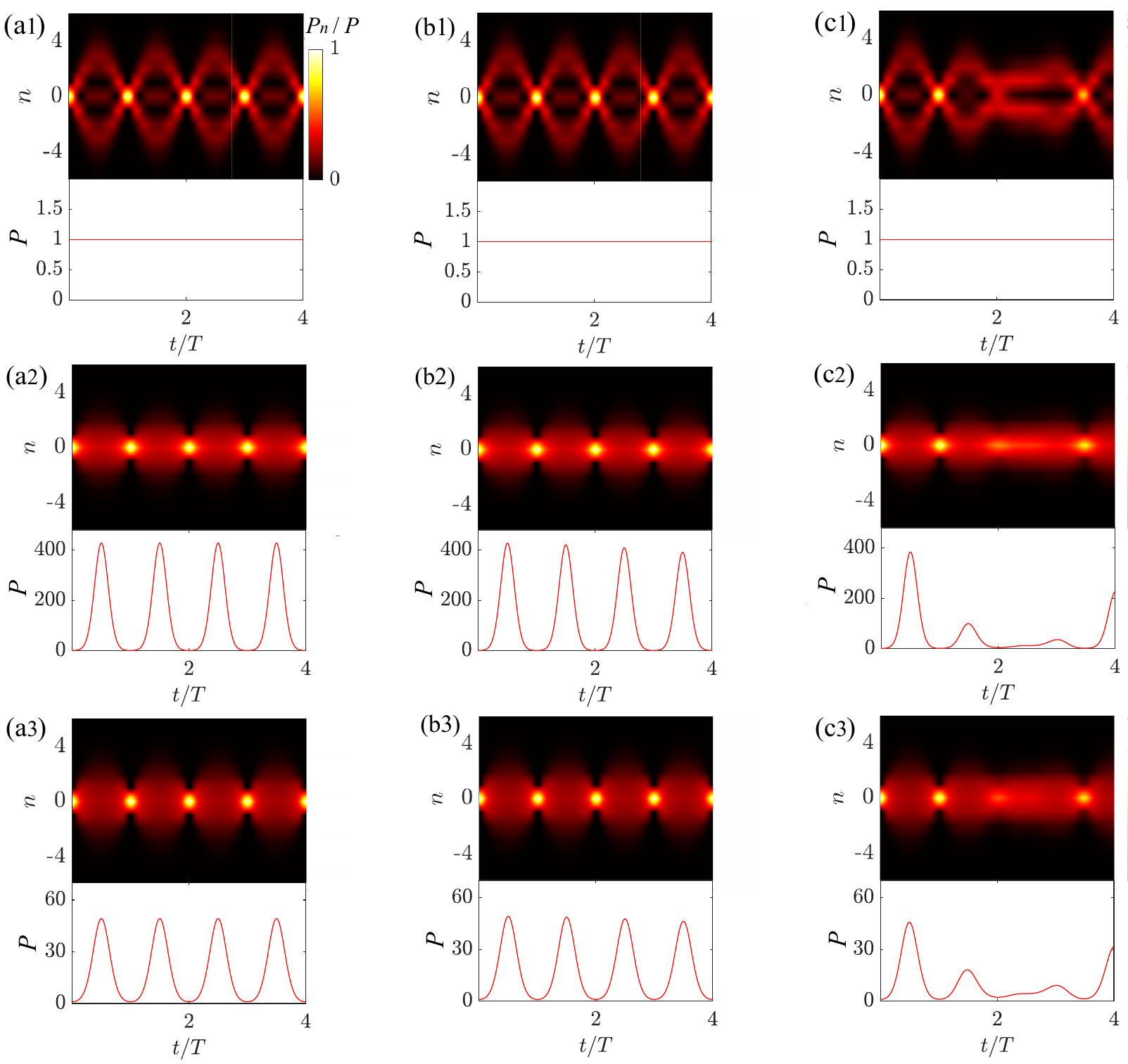}
\caption{{Plots of $P_{n}(t)/P(t)$ (see the main text)} and $P(t)$\ defined
in Eqs. (\protect\ref{Pm(t)} ) and (\protect\ref{P(t)}), obtained by
numerical diagonalization for several typical values of system parameters
with $\mathcal{\protect\omega }_{0}=1$. Here, (a1), (b1), and (c1)
correspond to the cases with $\protect\omega =0$, $0.01$, and $0.1$,
respectively, and $\protect\kappa _{0}=1$. (a2), (b2), and (c2) correspond
to the cases with $\protect\omega =0$, $0.01$ , and $0.1$, respectively, and 
$\protect\kappa _{0}=i$. (a3), (b3),and (c3) correspond to the cases with $%
\protect\omega =0$ , $0.01$, and $0.1$, respectively, and $\protect\kappa %
_{0}=e^{i\protect\pi /4}$. {We can see that the patterns in (a2, b2, c2) and
(a3, b3, c3) are similar. This inidcates that the imaginary part of hopping
plays an important role in the shapes of evolved states. }{According to the
analytical analysis, }${P(t)}${\ is periodic with the period depending on }$%
\protect\omega ${\ and }$\protect\omega _{0}$ {.\ When }$\protect\omega ${\
approaches }$\protect\omega _{0}$ {,\ the period becomes infinite. The
non-periodic behavior observed in (c1, c2, c3) arises because the scale
plotted in the figure is smaller than the period.}}
\label{fig2}
\end{figure}
\begin{figure}[tbh]
\centering
\includegraphics[width=0.93\textwidth]{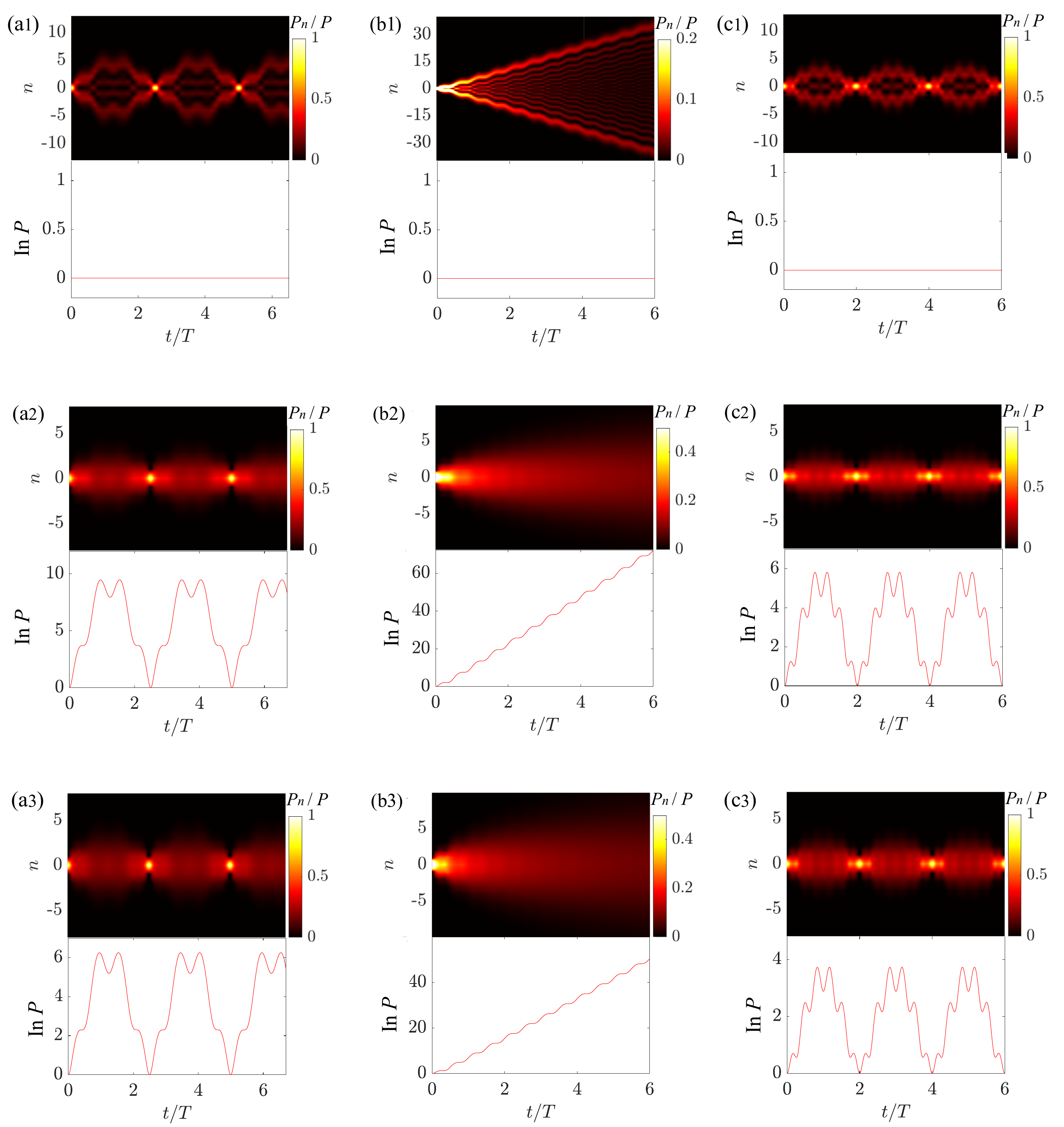}
\caption{{Plots of $P_{n}(t)/P(t)$ (see the main text) and $\text{In}P(t)$\
defined in Eqs. (\protect\ref{Pm(t)} ) and (\protect\ref{P(t)}), obtained by
numerical diagonalization for several typical values of system parameters
with $\mathcal{\protect\omega }_{0}=1$. Here, (a1), (b1), and (c1)
correspond to the cases with $\protect\omega =0.6$, $1$, and $1.5$,
respectively, and $\protect\kappa _{0}=1$. (a2), (b2), and (c2) correspond
to the cases with $\protect\omega =0.6$, $1$ , and $1.5$, respectively, and $%
\protect\kappa _{0}=i$. (a3), (b3),and (c3) correspond to the cases with $%
\protect\omega =0.6$, $1$, and $1.5$, respectively, and $\protect\kappa %
_{0}=e^{i\protect\pi /4}$.}}
\label{fig_plus}
\end{figure}

\section{Bloch oscillation in a non-Hermitian system}

\label{Bloch oscillation in non-Hermitian system} The dynamics of the system
are governed by the propagator 
\begin{equation}
U_{mn}\left( t\right) =\langle m|e^{-iHt}|n\rangle .
\end{equation}%
Employing the biorthonormal set $\left\{ |\varphi _{m}\rangle ,|\psi
_{n}\rangle \right\} $, we have 
\begin{eqnarray}
U_{mn}\left( t\right) &=&\sum_{l}\langle m|\psi _{l}\rangle
e^{-iE_{l}t}\langle \varphi _{l}|n\rangle  \notag \\
&=&J_{m-n}(-\frac{4\kappa }{\omega _{0}}\sin \frac{\omega _{0}t}{2}%
)e^{-i(m-n)\frac{\omega _{0}t}{2}}i^{m-n}e^{-in\omega _{0}t},  \notag \\
&=&i^{m-n}e^{-i(m+n)\frac{\omega _{0}t}{2}}J_{m-n}(-\frac{4\kappa }{\omega
_{0}}\sin \frac{\omega _{0}t}{2}),
\end{eqnarray}%
which is a periodic function with period $2\pi /\omega _{0}$. For a given
initial state 
\begin{equation}
\left\vert \Phi (0)\right\rangle =\sum_{l}c_{l}|l\rangle
\end{equation}%
the probability distribution in real space for the evolved state $\left\vert
\Phi (t)\right\rangle $ is 
\begin{equation}
P_{m}(t)=\left\vert \langle m\left\vert \Phi (t)\right\rangle \right\vert
^{2}=\left\vert \sum_{l}c_{l}U_{ml}\left( t\right) \right\vert ^{2},
\label{Pm(t)}
\end{equation}%
which is periodic. We obtain the conclusion that the total probability 
\begin{equation}
P(t)=\sum_{n}P_{n}(t),  \label{P(t)}
\end{equation}%
is still periodic, regardless of whether $\kappa $ is real or complex. {%
In practice, the measurement of $P_{n}(t)$ has been reported in
many non-Hermitian experiments \cite%
{Regensburger2012,StJean2017,Zhao2019,Liang2022}. }

{Up until now, we have been considering $\kappa(t)=\kappa$ as
time-independent; now, we explore what happens when $\kappa(t)$\ is time
dependent.} According to the adiabatic theorem, the expression of $%
U_{mn}\left( t\right) $\ still holds true for slowly varying $\kappa (t)$
during a short period of time. Considering the case with $\kappa (t)=\kappa
_{0}\cos (\omega t)$, where $\kappa _{0}$\ can be a complex number and the
frequency of $\kappa $\ is small, $\mathcal{\omega }\ll \mathcal{\omega }%
_{0} $. It can be predicted that $\left\vert \Phi (t)\right\rangle $ and $%
P(t)$\ are still approximately periodic even for the quasiadiabatic process.
However, this approach could be invalid for the diabatic process.

To verify and demonstrate the above analysis, numerical simulations are
performed to investigate the dynamic behavior of dynamic processes with
different values of $\mathcal{\omega }/\mathcal{\omega }_{0}$. We compute
the temporal evolution of a site state, $\left\vert \Phi (0)\right\rangle
=|0\rangle $ by using a uniform mesh in the time discretization for the
time-dependent Hamiltonian $H(t)$. The evolved state is 
\begin{equation}
\left\vert \Phi (t)\right\rangle =\mathcal{T}\exp \left[ -i\int_{0}^{t}H
\left( t\right) \mathrm{d}t\right] \left\vert \Phi (0)\right\rangle ,
\end{equation}
where $\mathcal{T}$\ is the time-order operator. {Figs. \ref{fig2}\ and \ref%
{fig_plus} are the plots of $P_{n}(t)/P(t)$ and $P(t)$ for several typical
values of $\omega /\omega _{0}$. For the non-Hermitian systems, the
difference between the maximum and minimum of $P(t)$\ is very large. Thus
the profile of the evolved state cannot be properly presented by the plot of 
$P_{n}(t)$. To present a complete and visible profile, we plot the rescaled $%
P_{n}$ by $P_{n}(t)/P(t)$. When the total probability $P(t)$ is periodic in
time, the periodicity of the evolved state can be confirmed by the profile
of the probability distribution. }

The numerical results indicate that the Bloch oscillation exists for small $%
\mathcal{\omega }/\mathcal{\omega }_{0}$. For non-Hermitian cases, the
profile of the evolved state does not change substantially, but the
probability is periodic. { From Fig. \ref{fig_plus} we can see
that the evolution of the profile of the evolved state is still periodic
when $\omega \mathbf{\neq }\omega _{0}$, but diffuses outward at $\omega
=\omega _{0}$. This indicates that the dynamics are peculiar at resonance,
which is the focus of the paper.}

\section{Floquet-resonance dynamics}

\label{Floquet-resonance dynamics}

Now, we investigate the dynamics of the system with time-dependent $\kappa
(t)$. {As previously mentioned, we consider periodically
modulated $\kappa (t)=\kappa _{0}\cos (\omega t)$, where $\kappa _{0}$\ can
be a complex number.} The solution of the Schr\"{o}dinger equation 
\begin{equation}
i\frac{\partial }{\partial t}|\Psi \left( t\right) \rangle =H(t)|\Psi
(t)\rangle
\end{equation}%
can be written in the form 
\begin{equation}
|\Psi (t)\rangle =\sum_{m}a_{m}(t)e^{-im\mathcal{\omega }_{0}t}|\psi
_{m}\rangle ,
\end{equation}%
where the coefficient $a_{n}$\ obeys the following equation: 
\begin{equation}
\frac{\partial a_{m}}{\partial t}=-\sum_{n}a_{n}e^{i\left( m-n\right) \omega
_{0}t}\left\langle \varphi _{m}\right\vert \frac{\partial }{\partial t}|\psi
_{n}\rangle .
\end{equation}%
Submitting the expression of $|\varphi _{m}(t)\rangle $ and $|\psi
_{n}(t)\rangle $, we have { 
\begin{eqnarray}
\frac{\partial a_{n}}{\partial t} &=&\frac{1}{\omega _{0}}\frac{\partial
\kappa }{\partial t}(a_{n-1}e^{i\mathcal{\omega }_{0}t}-a_{n+1}e^{-i\mathcal{%
\omega }_{0}t})  \notag \\
&=&-\frac{{\kappa _{0}}\omega }{\omega _{0}}{\sin (\omega t)}(a_{n-1}e^{i%
\mathcal{\omega }_{0}t}-a_{n+1}e^{-i\mathcal{\omega }_{0}t}),
\end{eqnarray}%
} which is the starting point for the following discussions.

Taking the rotating-wave approximation (RWA) under the condition $\left\vert 
\mathcal{\omega }- \mathcal{\omega }_{0}\right\vert \ll \mathcal{\omega }%
_{0} $, we have 
\begin{equation}
i\frac{\partial a_{n}}{\partial t}=\frac{\kappa _{0}\omega }{2\omega _{0}}
[a_{n-1}e^{i(\mathcal{\omega }_{0}-\omega )t}+a_{n+1}e^{i(\omega -\mathcal{\
\omega }_{0})t}].
\end{equation}
We focus on the dynamics of the system at resonance $\mathcal{\omega }= 
\mathcal{\omega }_{0}$, in which the equation becomes 
\begin{equation}
i\frac{\partial a_{n}}{\partial t}=\frac{\kappa _{0}}{2}(a_{n-1}+a_{n+1}).
\end{equation}
The Schr\"odinger equation is used for a uniform tight-binding chain.

lt clearly shows that the eigenvalues of the equivalent Hamiltonian are
proportional to the hopping constant $\kappa_{0}$, and then determines the
behavior of the amplitude of the evolved state which can be conservative,
damping, or explosive, respectively. The real and imaginary parts of $%
\kappa_{0}$ will affect the real and imaginary nature of the energy levels,
which in turn will affect whether the probability is conserved and the
evolution of the wave function. It can be anticipated that the presence of
an imaginary part in $\kappa_{0}$ will significantly influence the dynamical
evolution of the system.

\section{Dynamical exponents}

\label{Dynamical exponents}

As shown above, the resonant dynamics are essentially governed by the simple
equivalent Hamiltonian 
\begin{equation}
H_{\mathrm{eq}}=\frac{\kappa _{0}}{2}\sum\limits_{n=-\infty }^{+\infty
}(|n\rangle \langle n+1|+\text{\textrm{H.c}}),
\end{equation}%
where $|n\rangle $\ denotes the eigenstates of the instantaneous
Hamiltonian. {{The dynamics of }$H_{\mathrm{eq}}$\ were investigated in Ref. 
\cite{Longhi2017}, which indicates that the reality of $\kappa _{0}$\ plays
an important role in the dynamics.} The Hamiltonian $H_{\mathrm{eq}}$\ can
be diagonalized in the form 
\begin{equation}
H_{\mathrm{eq}}=\sum\limits_{k\in (-\pi ,\pi )}\varepsilon _{k}|k\rangle
\langle k|,
\end{equation}%
with 
\begin{equation}
|k\rangle =\frac{1}{\sqrt{2\pi }}\sum\limits_{l}e^{ikl}|l\rangle ,
\end{equation}%
and the spectrum 
\begin{equation}
\varepsilon _{k}=\kappa _{0}\cos k.
\end{equation}%
The time evolution of state 
\begin{equation}
\left\vert \Psi (t)\right\rangle =\sum\limits_{l}a_{l}(t)|l\rangle
=\sum\limits_{l}a_{l}(0)\sum_{n}i^{n-l}J_{n-l}(-\kappa _{0}t)|n\rangle ,
\end{equation}%
can be obtained for any given $\left\{ a_{l}(0)\right\} $. Intuitively, an
initial state $|0\rangle $\ (or $a_{l}(0)=\delta _{0l}$) could spread out
and never return whether $\kappa _{0}$ is real or complex. In the following,
we focus on two cases with $\kappa _{0}=1$\ and $\kappa _{0}=i$.

\begin{figure*}[t]
\centering
\includegraphics[width=0.92\textwidth]{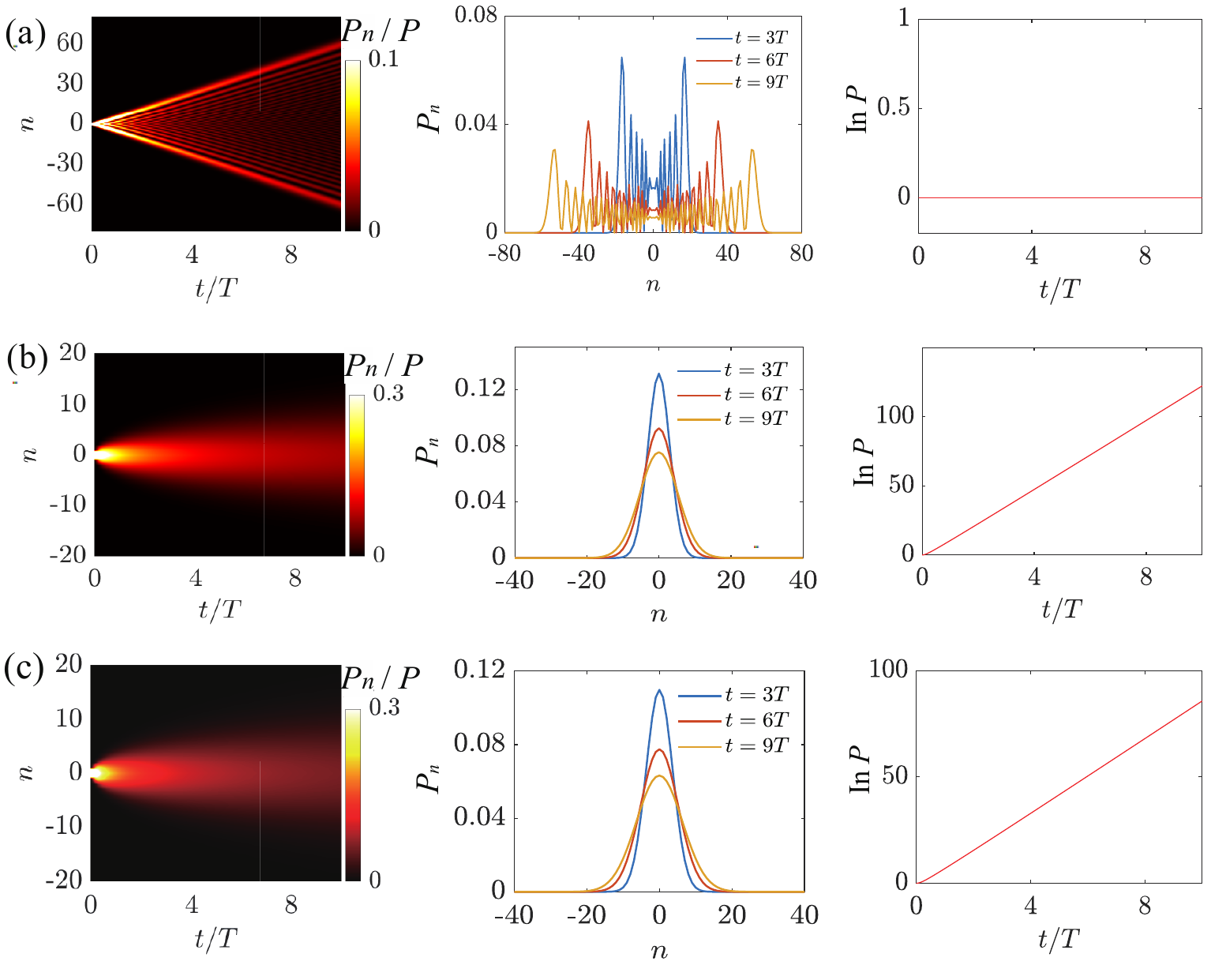}
\caption{The same plots of {$P_{n}(t)/P(t)$} and $\text{In}P(t)$\ as those
in Fig. \protect\ref{fig2} are shown for several typical values of system
parameters but at a resonance $\mathcal{\protect\omega }_{0}=\protect\omega %
=1$. Here, (a), (b), and (c) correspond to the cases with $\protect\kappa %
_{0}=1$, $i$, and $e^{i\protect\pi /4}$, respectively. The middle panels
correspond to the probability distributions at moments $t=3T$, $6T$, and $9T$
, respectively. The right panels are the logarithmic plots of total
probabilities as functions of time. These results agree with our analytical
analysis in the main text. {We can see that the patterns in (b) and (c) are
almost identical. This inidcates that the imaginary part of hopping plays
the important role in the energy-level spreading.} }
\label{fig3}
\end{figure*}

(i) $\kappa _{0}=1$; in this case, the straightforward derivation yields 
\begin{equation}
a_{n}(t)=i^{n}J_{n}(-t),
\end{equation}%
which ensures the probability distribution of the energy levels 
\begin{equation}
\mathcal{P}_{n}(t)=\left\vert a_{n}(t)\right\vert ^{2}=\left\vert
J_{n}(-t)\right\vert ^{2},
\end{equation}%
and 
\begin{equation}
\mathcal{P}=\sum\limits_{n}\mathcal{P}_{n}(t)=1.
\end{equation}%
The feature of the Bessel function indicates that there exists a maximum at
the edge of $\mathcal{P}_{n}(t)$, which can be regarded as the wave front of
the occupied-energy-level spreading. The location of such a wave front can
then be determined by the equation 
\begin{equation}
{\frac{\partial \mathcal{P}_{n_{c}}(t)}{\partial t}=0.}
\end{equation}%
Based on the relations of Bessel functions 
\begin{equation}
J_{n_{c}-1}(x)-J_{n_{c}+1}(x)=2J_{n_{c}}^{\prime
}(x),\,\,\,J_{n_{c}+1}\left( x\right) +J_{n_{c}-1}\left( x\right)
=2\left\vert n_{c}/t\right\vert J_{n_{c}}\left( x\right) ,
\end{equation}%
we have 
\begin{equation}
{2J_{n_{c}+1}(-t)=2|n_{c}/t|J_{n_{c}}(-t).}
\end{equation}%
For large $n$ and $t$, we have ${J_{n_{c}+1}(-t)\approx J_{n_{c}}(-t)}$,
which results in 
\begin{equation}
{|n_{c}|\approx t.}
\end{equation}%
Then, we conclude that the dynamical exponent of occupied-energy-level
spreading\ is $z=1$.

(ii) $\kappa _{0}=i$; in parallel, the corresponding amplitude is the Bessel
function with imaginary argument 
\begin{equation}
a_{2n}(t)=(-1)^{n}J_{2n}(-it)=\frac{(-1)^{n}}{2\pi }\int_{-\pi }^{\pi }\cos
(2nk)e^{-t\sin k}\mathrm{d}k,
\end{equation}%
and 
\begin{equation}
a_{2n+1}(t)=i(-1)^{n}J_{2n+1}(-it)=\frac{(-1)^{n+1}}{2\pi i}\int_{-\pi
}^{\pi }\sin [(2n+1)k]e^{-t\sin k}\mathrm{d}k.
\end{equation}%
On a long time scale, the main contribution to the integral comes from the
integrand around $k=-\pi /2$. Then, we have 
\begin{equation}
a_{2n}(t)\approx e^{-\frac{(2n)^{2}}{2t}}\frac{1}{2\pi }\int_{-\pi }^{\pi
}e^{-t\sin k}\mathrm{d}k,
\end{equation}%
and 
\begin{equation}
a_{2n+1}(t)\approx e^{-\frac{(2n+1)^{2}}{2t}}\frac{1}{2\pi i}\int_{-\pi
}^{\pi }e^{-t\sin k}\mathrm{d}k,
\end{equation}%
with small $\varepsilon _{k}$, which results in the probability distribution
of the energy levels 
\begin{equation}
\mathcal{P}_{n}(t)=|a_{n}(t)|^{2}\approx \sqrt{\frac{1}{\pi t}}\mathcal{P}%
e^{-\frac{n^{2}}{t}},
\end{equation}%
with the total probability 
\begin{equation}
\mathcal{P}=\sum_{n}\mathcal{P}_{n}=\frac{1}{2\pi }\int_{-\pi }^{\pi
}e^{-2t\sin k}\mathrm{d}k.
\end{equation}%
Obviously, the profile of $\mathcal{P}_{n}$\ is a Gaussian type, and the
total probability grows exponentially as $e^{2t}$ approaches the large $t$
limit. In this situation, the dynamics of occupied-energy-level spreading
can be characterized by the full width at half maximum, 
\begin{equation}
\mathcal{P}_{n_{c}}(t)=\frac{1}{2}\mathcal{P}_{0}(t),
\end{equation}%
which results in 
\begin{equation}
|n_{c}|\approx \sqrt{t}.
\end{equation}%
Then, we conclude that the dynamical exponent of occupied-energy-level
spreading\ is $z=1/2$.

{The appendix \ref{Appendix} presents the dynamical exponent for $%
\kappa_0$ as any complex number. When it is real, $z=1$; when it is not
real, $z=1/2$. The non-Hermitian case differs from the Hermitian case
primarily due to the imaginary part of $\kappa_0$.} To verify and
demonstrate the above analysis, numerical simulations are performed to
investigate the dynamic behavior of dynamic processes at resonance with $%
\mathcal{\omega }=\mathcal{\omega }_{0}$. For the same initial state as
shown in Fig. \ref{fig2}, the plots of $P_{n}(t)$ and $P(t) $\ for several
typical values of $\kappa _{0}$ \ are presented in Fig. \ref{fig3}. These
numerical results agree with our above analysis in two aspects: (i) At
resonance, the dynamics are no longer periodic; (ii) The level spreading
exhibits distinct behaviors for Hermitian and non-Hermitian systems.

\section{2D simulator via wavepacket dynamics}

\label{2D simulator via wavepacket dynamics}

In this section, we will show how the time evolution of a wavepacket in an
engineered time-independent 2D square lattice can simulate our results for
Floquet dynamics in a 1D time-dependent lattice. In experiments,
single-particle hopping dynamics can be simulated by discretized spatial
light transport in an engineered 2D square lattice of evanescently coupled
optical waveguides \cite{Christodoulides2003}. A 2D lattice can be
fabricated by coupled waveguides, by which the temporal evolution of the
single-particle probability distribution\ in the 2D lattice can be
visualized by the spatial propagation of the light intensity.

We consider a single-particle Hamiltonian on a square lattice 
\begin{eqnarray}
H_{\mathrm{2D}} &=&\kappa _{0}\sum_{n,m=-\infty }^{\infty }\cos \left(
qm\right) |n,m\rangle \langle n+1,m| -J\sum_{n,m=-\infty }^{\infty
}|n,m\rangle \langle n,m+1|+\mathrm{h.c.}  \notag \\
&&+\omega _{0}\sum_{n,m=-\infty }^{\infty }n|n,m\rangle \langle n,m|,
\label{H2D}
\end{eqnarray}
which is an array of coupled 1D Bloch systems with $m$-dependent hopping
strength $\kappa _{0}\cos \left( qm\right) $. Fig. \ref{fig1}(b) shows a
schematic of the system. {According to the above analysis, the
system is governed by Bloch dynamics with frequency $\omega _{0}$ for
vanishing $J$, regardless of whether $\kappa _{0}$\ is real or complex.}

\begin{figure}[h]
\centering
\includegraphics[width=0.9\textwidth]{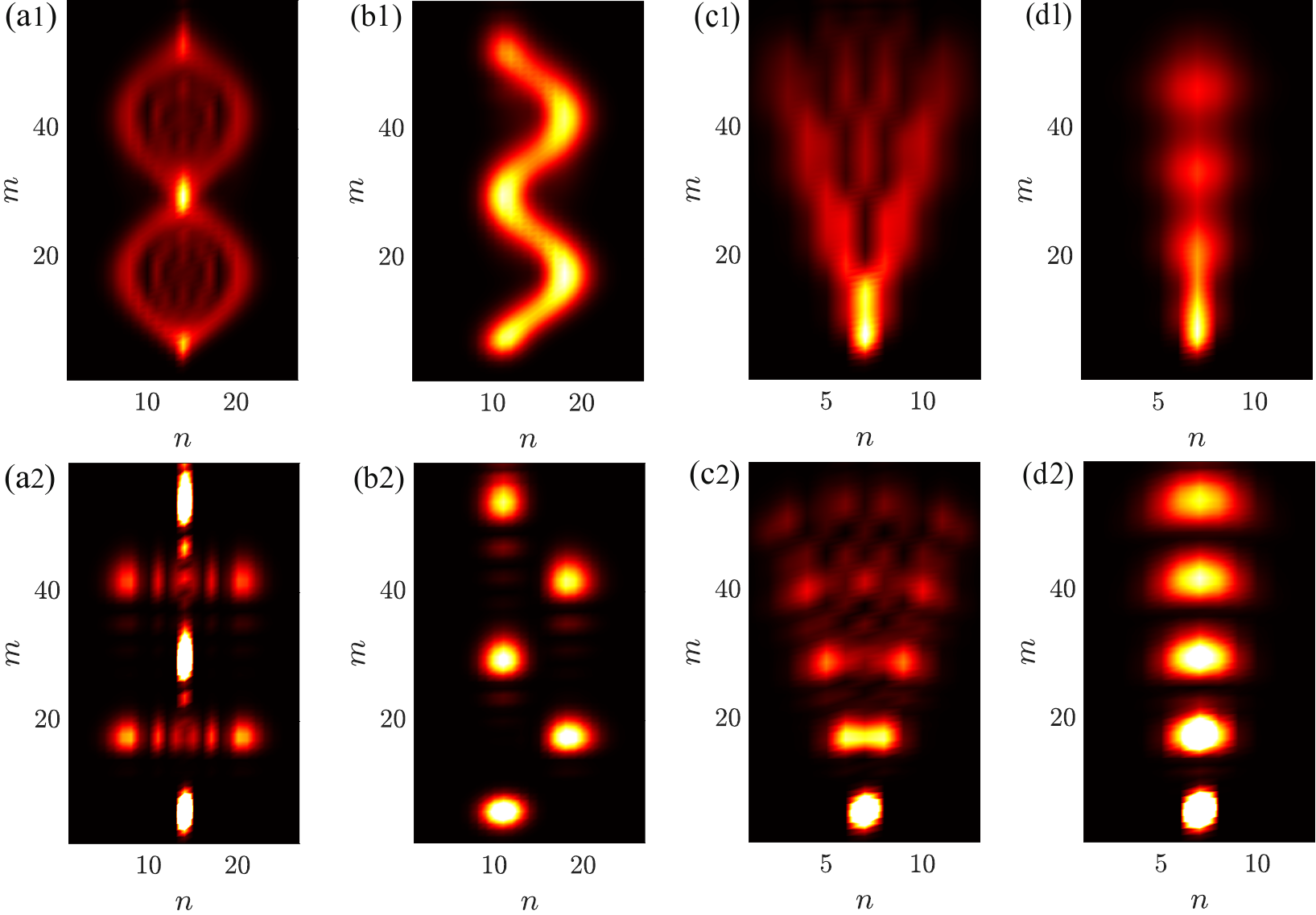}
\caption{For numerical simulations of the dynamics, the trajectories of
wavepackets evolving in the in a 2D lattice with four typical configurations
were obtained. The traces of the wavepacket $P(n,m)$ for the time evolutions
of the four cases are shown in (a1), (b1), (c1), and (d1). The probability
distributions $p(n,m,t)$ for the time evolutions of four representative
cases\ at $t=0$, $t=2\protect\pi /J$, $t=4\protect\pi /J$, $t=6\protect\pi %
/J $ and $t=8\protect\pi /J\,$are shown in (a2), (b2), (c2), and (d2),
respectively. The initial states and system parameters are given in the main
text. The size of the system is $60\times 30$. The energy scale of the
Hamiltonian is taken as $J=1 $. The probability distributions on the square
lattice exhibit distinct patterns corresponding to the dynamic behaviors of
Bloch breathing, oscillation, diffusion and superdiffusion.}
\label{fig4}
\end{figure}

Now, we start to investigate the dynamics in the presence of nonzero $J$.
The $k$-space representation of the Schr\"{o}dinger equation 
\begin{equation}
i\frac{\partial }{\partial t}|\psi \rangle =H_{\mathrm{2D}}|\psi \rangle ,
\end{equation}%
can be written as 
\begin{eqnarray}
i\frac{\partial \psi }{\partial t} &=&\kappa _{0}\cos k_{x}[\psi \left(
k_{x},k_{y}+q,t\right) +\psi \left( k_{x},k_{y}-q,t\right) ]  \notag \\
&&+i\omega _{0}\frac{\partial \psi }{\partial k_{x}}-2J\cos k_{y}\psi \left(
k_{x},k_{y},t\right) ,
\end{eqnarray}%
by taking the Fourier transformation 
\begin{equation}
|k_{x},k_{y}\rangle =\frac{1}{2\pi }\sum_{n,m=-\infty }^{\infty }e^{i\left(
k_{x}n+k_{y}m\right) }|n,m\rangle .
\end{equation}%
For the case with $q=0$, the solution $\psi \left( k_{x},k_{y},t\right) $
reduces to 
\begin{equation}
\psi \left( k_{x},k_{y},t\right) =f\left( k_{x},t\right) e^{it2J\cos k_{y}},
\end{equation}%
which resembles that in Ref.\ \cite{Mallick2021}, where $f\left(
k_{x},t\right) $\ obeys the following equation: 
\begin{equation}
i\frac{\partial f}{\partial t}=(2\kappa _{0}\cos k_{x}+i\omega _{0}\frac{%
\partial }{\partial k_{x}})f\left( k_{x},t\right) .
\end{equation}%
In the case with $k_{y}\approx \pi /2$, we have $\cos k_{y}\approx k_{y}-\pi
/2$. One can construct a solution by the superposition of states $%
|k_{x},k_{y}\rangle $\ with $k_{y}\approx \pi /2$ in the form \cite{Kim2006} 
\begin{eqnarray}
\left\vert g\left( k_{x},t\right) \right\rangle  &=&(\frac{2\alpha ^{2}}{\pi 
})^{1/4}\int_{-\pi }^{\pi }e^{-\alpha ^{2}\left( k_{y}-\pi /2\right)
^{2}}e^{it2J\cos k_{y}}f\left( k_{x},t\right) \mathrm{d}k_{y}|k_{x},k_{y}%
\rangle   \notag \\
&\approx &(\frac{1}{2\pi \alpha ^{2}})^{1/4}\sum_{m}e^{im\pi
/2}e^{-(1/4)(m-2Jt)^{2}/\alpha ^{2}}f\left( k_{x},t\right) |k_{x},m\rangle ,
\end{eqnarray}%
which is a traveling Gaussian wavepacket in the $y$ direction. It is
presumed that such a solution still holds true by taking $\kappa
_{0}\rightarrow \kappa _{0}\cos \left( qm\right) $ when $\kappa _{0}\cos
\left( qm\right) $\ varies slowly during the extension of the wavepacket in
the y direction due to the locality of the solution. In this sense, the
traveling wavepacket experiences an adiabatic process, after which the
independence of the dynamics in both directions is maintained. Obviously, a
small value of $q$ should be needed. On the other hand, the value of $\omega
_{0}$ must match $q$\ when the system is employed to simulate the dynamics
at the resonance. To this end, a small value of {$|\kappa _{0}|$}
should be taken to maintain the Stark ladder.

Actually, under the conditions { $|\kappa _{0}|\ll J$ }and $%
k_{y}\approx \pi /2$, such a 2D system is equivalent to a coupled waveguide
array due to the linear dispersion in the $y$ direction, in which the
tunneling rate between neighboring waveguides is $m$ dependent. The speed of
the propagating wavepacket along $m$ is $2J$. To simulate the dynamics along 
$x$, one can simply take $q=\omega /\left( 2J\right) $, regarding $m$\ as
the temporal coordinate.

{To simulate the time-dependent evolution of a 1D system (\ref{e1}%
) in a 2D lattice system (\ref{H2D}), we design specific initial states 
\begin{equation}
\psi (n,m,0)=x(n)y(m),
\end{equation}%
where $x(n)$ is taken as the initial state of the 1D system and $y(m)=({4\pi
^{2}})^{-1/4}e^{im\pi /2}e^{-m^{2}/4}$ represents a Gaussian wave packet
with momentum ${\pi }/{2}$ in the y-direction, serving as an alternative to
the time axis. At first, we verify this scheme by considering the Bloch
oscillation in Hermitian system. (i) To simulate Bloch breathing, $x(n)$ is
taken as $\delta _{n,n_{0}}$. The system parameters are $\kappa _{0}=J$, $%
\omega =0$ and $\omega _{0}=0.5J$. (ii) To simulate Bloch oscillation of a
wavepacket, $x(n)$ is taken as $e^{-{(n-n_{0})^{2}}/4}$. The system
parameters are $\kappa _{0}=J$, $\omega =0$,\ and $\omega _{0}=0.5J$.
Second, we consider the simulations for the dynamics, which are presented in
Fig. \ref{fig_plus}(b1) and (b2). (iii) To simulate superdiffusion, $x(n)$ is taken as $%
\delta _{n,n_{0}}$. The system parameters are $\kappa _{0}=0.25J$ and $%
\omega =\omega _{0}=0.5J$. (iv) To simulate diffusion, $x(n)$ is taken as $%
\delta _{n,n_{0}}$. The system parameters are $\kappa _{0}=0.25Ji$ and $%
\omega =\omega _{0}=0.5J$. {For greater clarity, the specific parameters are
shown in Tab. \ref{tab1}.} The evolved state has the form 
\begin{equation}
\left\vert \psi \left( n,m,t\right) \right\rangle =e^{-iH_{\mathrm{2D}%
}t}\left\vert \psi \left( n,m,0\right) \right\rangle ,
\end{equation}%
which can be computed by exact diagonalization.}

\begin{table}[tb]
\caption{{List of the parameters of the
four
driven
Hamiltonians
and initial
states,
where
\(
y(m)=({4\protect\pi
^{2}})^{-1/4}e^{im\protect\pi
/2}e^{-m^{2}/4}
\)
represents a Gaussian wave
packet with momentum
\(
{\pi}/{2} \) in the
y-direction, moving at a
constant velocity \( v =
2J \) in
the
y-direction. Therefore, the distance
the wave packet travels
in
the
y-direction represents time.} }
\label{tab1}\centering
\textbf{\renewcommand\arraystretch{1}} \renewcommand{\arraystretch}{1.3} %
\setlength{\tabcolsep}{22pt} 
\par
\begin{tabular}{ccccl}
\hline\hline
& $\kappa_0$ & $\omega$ & $\omega_0$ & Initial state \\ \hline
(i) & $J$ & 0 & 0.5$J$ & ${\delta_{n, n_{0}}y(m)} $ \\ 
(ii) & $J$ & 0 & 0.5$J$ & ${e^{-{(n-n_0)^2}/4}y(m)}$ \\ 
(iii) & 0.25$J$ & 0.5$J$ & 0.5$J$ & $\delta_{n, n_{0}}y(m)$ \\ 
(iv) & 0.25$Ji$ & 0.5$J$ & 0.5$J$ & $\delta_{n, n_{0}}y(m)$ \\ \hline\hline
\end{tabular}
\end{table}

The probability distribution at position $(n,m)$ at time $t$ is defined by
the probabilities 
\begin{equation}
p(n,m,t)=\left\vert \langle n,m,\left\vert \psi \left( n,m,t\right)
\right\rangle \right\vert ^{2}.  \label{p_nmt}
\end{equation}
To visualize the profile of the dynamics, we define the function 
\begin{equation}
P(n,m)=\sum_{j=1}p(n,m,j\tau ),  \label{trace}
\end{equation}
to record the trace of the wavepacket. Here, we use $j\tau $\ to denote the
discretized time coordinate. It is expected that the distribution of $P(n,m)$
\ exhibits patterns similar to those in Fig. \ref{fig3}. {The
numerical results presented in Fig. \ref{fig4}\ show that the probability
distributions on the square lattice exhibit distinct patterns corresponding
to the dynamic behaviors of Bloch breathing, oscillation, superdiffusion and
diffusion.}

\section{Conclusion}

\label{Conclusion}

In summary, we simultaneously examine the effects of non-Hermiticity and
periodic driving on a Stark ladder system. The most fascinating and
important feature of such systems is that the Stark ladder can be maintained
regardless of the non-Hermiticity of the hopping strength, and the energy
level spacing depends only on its norm. This allows us to solve the problem
in an analytical manner. The Bloch oscillation is broken when the Floquet
frequency is resonant with the energy level spacing, and the occupied energy
levels spread out. Furthermore, analytic analysis and numerical simulation
show that the level-spreading dynamics for real and complex hopping
strengths exhibit distinct behaviors of superdiffusion and diffusion,
respectively. In addition, we propose a scheme to simulate the results by
wavepacket dynamics in a 2D square lattice. This can be experimentally
realized in the coupled array of waveguides. These findings may help us
deepen our understanding of the interplay between non-Hermiticity and
periodic driving impacts on the dynamics of the system.

\section*{Acknowledgements}

\paragraph{Funding information}

This work was supported by the National Natural Science Foundation of China
(under Grant No. 12374461).

\section*{Appendix}

\label{Appendix}

{In this appendix \ref{Appendix}, we derive the dynamical
exponent for the case in which $\kappa _{0}$ is an arbitrary complex number.
When $\kappa _{0}$ is a real number, i.e., $\text{Im}(\kappa _{0})=0$, using
the same Bessel function estimation method as in the main text, one can
obtain the dynamical exponent of occupied-energy-level spreading $z=1$, and
the total probability $\mathcal{P}$ remains constant at 1. 

When $\text{Im}(\kappa _{0})\neq 0$, we have%
\begin{equation}
a_{2n}(t)=(-1)^{n}J_{2n}(-\kappa _{0}t)=\frac{(-1)^{n}}{2\pi }\int_{-\pi
}^{\pi }\cos (2nk)e^{i\kappa _{0}t\sin k}\mathrm{d}k,
\end{equation}%
and 
\begin{equation}
a_{2n+1}(t)=i(-1)^{n}J_{2n+1}(-\kappa _{0}t)=\frac{(-1)^{n+1}}{2\pi i}%
\int_{-\pi }^{\pi }\sin [(2n+1)k]e^{i\kappa _{0}t\sin k}\mathrm{d}k.
\end{equation}%
On a long time scale, the main contribution to the integral comes from the
integrand around $k=-\pi /2$(or $\pi /2$, depends on the sign of $\text{Im}%
(\kappa _{0})$). Then, we have 
\begin{eqnarray}
a_{2n}(t) &\approx &{e^{-\frac{(2n)^{2}}{2i\kappa _{0}t}}\frac{1}{2\pi }%
\int_{-\pi }^{\pi }e^{i\kappa _{0}t\sin k}\mathrm{d}k,\,\text{Im}(\kappa
_{0})<0} \\
&\approx &{e^{\frac{(2n)^{2}}{2i\kappa _{0}t}}\frac{1}{2\pi }\int_{-\pi
}^{\pi }e^{i\kappa _{0}t\sin k}\mathrm{d}k,\,\text{Im}(\kappa _{0})>0}
\end{eqnarray}%
and 
\begin{eqnarray}
a_{2n+1}(t) &\approx &{e^{-\frac{(2n+1)^{2}}{2i\kappa _{0}t}}\frac{i}{2\pi }%
\int_{-\pi }^{\pi }e^{i\kappa _{0}t\sin k}\mathrm{d}k,\,\text{Im}(\kappa
_{0})<0} \\
&\approx &{e^{\frac{(2n+1)^{2}}{2i\kappa _{0}t}}\frac{1}{2\pi i}\int_{-\pi
}^{\pi }e^{i\kappa _{0}t\sin k}\mathrm{d}k,\,\text{Im}(\kappa _{0})>0}
\end{eqnarray}%
with small $\varepsilon _{k}$, which results in the probability distribution
of the energy levels 
\begin{equation}
\mathcal{P}_{n}(t)=|a_{n}(t)|^{2}\approx \sqrt{\frac{|\text{Im}(\kappa _{0})|%
}{\pi |\kappa _{0}|^{2}t}}\mathcal{P}e^{-\frac{|\text{Im}(\kappa _{0})|n^{2}%
}{|\kappa _{0}|^{2}t}},
\end{equation}%
with the total probability 
\begin{equation}
\mathcal{P}=\sum_{n}\mathcal{P}_{n}=\frac{1}{2\pi }\int_{-\pi }^{\pi }e^{-2%
\text{Im}(\kappa _{0})t\sin k}\mathrm{d}k.
\end{equation}%
Obviously, the profile of $\mathcal{P}_{n}$\ is a Gaussian type, and the
total probability grows exponentially as $e^{2|\text{Im}(\kappa _{0})|t}$
approaches the large $t$ limit. In this situation, the dynamics of
occupied-energy-level spreading can be characterized by the full width at
half maximum, 
\begin{equation}
\mathcal{P}_{n_{c}}(t)=\frac{1}{2}\mathcal{P}_{0}(t),
\end{equation}%
which results in 
\begin{equation}
|n_{c}|\approx \sqrt{t}.
\end{equation}%
Then, we conclude that the dynamical exponent of occupied-energy-level
spreading\ is $z=1/2$.}

\providecommand{\newblock}{}

\end{document}